\documentclass[10pt, hidelinks]{article}

\usepackage[nocrop,newcenter,frame]{pagedims}

\usepackage{epsfig}

\usepackage{graphicx}

\usepackage{stata}

\usepackage{amsmath}

\usepackage{shadow}

\usepackage{threeparttable} 
\usepackage{tabularx} 

\usepackage{authblk}
\usepackage{hyperref}
\usepackage{natbib}
\usepackage{supertabular}

\newcommand{\mono}[1]{\texttt{#1}}

\begin{document}
\title{\texttt{require}: Package dependencies for reproducible research
\thanks{We thank Sebastian Kranz for helping us access the Stata code contained in the replication packages published by the \textit{American Economic Association} journals, \textit{the Review of Economic Studies}, and \textit{the Review of Economics and Statistics}. We are also grateful to Stephen P. Jenkins (editor), Benjamin Daniels, Paulo Guimarães, Miklós Koren, Julian Reif, Luis Eduardo San Martin, Lars Vilhuber, and seminar participants at the 2023 Stata Conference for their valuable suggestions.}
}

\author[1]{Sergio Correia}
\author[2]{Matthew P. Seay}
\affil[1]{Board of Governors of the Federal Reserve System, \href{mailto:sergio.a.correia@frb.gov}{sergio.a.correia@frb.gov}}
\affil[2]{Board of Governors of the Federal Reserve System, \href{mailto:matt.seay@frb.gov}{matt.seay@frb.gov}}
\maketitle

\begin{abstract}
The ability to conduct reproducible research in Stata is often limited by the lack of version control for community-contributed packages.
This article introduces the \texttt{require} command, a tool designed to ensure Stata package dependencies are compatible across users and computer systems.
Given a list of Stata packages, \texttt{require} verifies that each package is installed,  checks for a minimum or exact version or package release date, and optionally installs the package if prompted by the researcher.


\end{abstract}

Keywords:
\mono{require}, \mono{ssc}, \mono{which}, reproducible research, package management, dependency management, GitHub

\section{Introduction}

In this article, we introduce the \stcmd{require} command, which provides a simple way of managing Stata package dependencies and ensuring their compatibility.
By using \stcmd{require}, researchers can increase the reproducibility of their work by specifying the version requirements of any package dependencies used, and ensuring that their code will only run if these requirements are met.

Reproducibility is the ability to create identical results of a prior study, given the same set of tools, code, and data \citep{bollen2015, turing}.
Despite the introduction of data and code availability standards by leading academic journals \citep{dcas} as well as many coding improvements such as version control, reproducibility studies consistently show that a material share of published research papers are not reproducible.
For instance, \citet{Vilhuber2021} found that 62 percent of the research articles published in \textit{American Economic Journal: Applied Economics} between 2009 and 2018 were not reproducible.
Although there are numerous challenges to creating reproducible research, a common obstacle is inconsistent output caused by users employing different versions of community-contributed packages.\footnote{Community-contributed packages are commands or collections of commands produced and maintained by the Stata community. They comprise a vast list of topics, ranging from output generation (\stcmd{estout}, \stcmd{outreg2}) and data manipulation (\stcmd{gtools}) to advanced regression routines (\stcmd{ivreg2}, \stcmd{rdrobust}) and more. Historically, these have been usually distributed through the Boston College Statistical Software Components (SSC) archive \citep{cox2005}. However, in recent times, platforms such as GitHub and GitLab, which are based on distributed version control systems, have also become popular distribution channels.}

Package version control remains a challenge for Stata users for at least a few reasons.
First, as community-contributed packages get updated over time, newer versions might introduce slight changes in their output.
This makes research results less reproducible unless researchers document the exact package versions used and these versions remain broadly available to others.
This issue has become even more salient with the recent increase in popularity of distributed version control systems, such as Git, as software development now prioritizes shorter release cycles with a higher frequency of version changes.
For instance, as of April 2024, the \stcmd{gtools} package maintained on GitHub has accumulated 709 incremental updates in the form of "commits" and 19 substantial updates published "releases".



For more concrete examples, consider the following situations where the absence of package version control might compromise the reproducibility and even the quality of research:

\begin{enumerate}
    \item When working across multiple computers: Researchers often use file synchronization services---such as Dropbox, OneDrive, and Google Drive---to backup and sync their code and data. This practice enables them to easily switch between laptop and desktop computers or to reinstall or change their operating system. This workflow, however, might lead to errors unless researchers can ensure consistent package versions across their devices.
    \item In secure research environments: Package version updates are often pushed silently to users by network administrators. These updates can lead to subtle changes in results that may be difficult detect without a package version control system.
    \item In co-authored research: Different co-authors may inadvertently use inconsistent versions of regression packages, like \stcmd{ivreg2}, \stcmd{rdrobust}, or \stcmd{reghdfe}. Such inconsistencies can lead to slightly different point estimates and standard errors, which are often challenging to diagnose. 
    \item During the peer-review process: Research papers can take years from inception to publication. Utilizing explicitly-defined package versions can help maintain the consistency of output across review rounds, potentially streamlining the review process.
    \item In evaluations by journal data editors: Many journals now mandate authors to distribute both their code and data prior to publication \citep{dcas}.
    However, replication materials often fail to execute without code modifications, typically due to the usage of absolute file paths and, more relevantly, missing package dependencies. Clearly specifying package dependencies in the code can improve the quality of journal replication materials and increase their utility for other researchers. Further, this can also alleviate the workload of journal data editors, encouraging more journals to adopt code and data availability policies \citep{Vilhuber2021}.
\end{enumerate}

While tools to manage package dependencies are available in other languages like Python (\texttt{requirements.txt}) and R (\texttt{packrat}), Stata currently lacks this feature. To bridge this gap, we introduce the \stcmd{require} package. \stcmd{require} enables users to explicitly declare package dependencies along with their minimum or exact versions, verify if these requirements are met, and, if asked by the user, install the packages that are missing or have an incompatible version.

Within the Stata ecosystem, \stcmd{require} complements other packages such as \stcmd{github} \citep{haghish2020} which manages GitHub-based packages, \stcmd{repkit} \citep{repkit}, which sets project-specific ado paths, and \stcmd{dependencies} \citep{goldemberg2019}, which saves a user's ado path into a zipfile so it can be shared or restored at a later date.\footnote{Note that redistributing ado paths has limitations. For example, when dealing with compiled packages, Stata only downloads and installs the components specific to the user's operating system. Thus, an ado-path zipped on e.g. a Windows installation might not work on Linux or macOS environments.}


In order to fully benefit from using \stcmd{require}, we encourage researchers to pair it with other reproducibility tools and best practices.
For a detailed list of best practices, see  \citet{repro_tips_andrade, repro_tips_baum, repro_tips_guimaraes, mcway2021, reif}.
Additionally, for tools specializing in reproducible report generation, consider \citet{repro_doc_rising, repro_doc_baldridge, repro_doc_haghish, repro_doc_macdonald, repro_doc_peng}.

In Section 2, we describe how  \stcmd{require} works, including its syntax, structure, and basic usage. Section 3 contains a few examples of the command and best practices for using it in research projects. In Sections 4 and 5 we discuss potential limitations of \stcmd{require}, how it performs when tested against packages that researchers rely on most. Section 6 concludes and proposes a few areas for future work. 
\newcommand{\Ret}[2]{\stcmd{s(#1)} & #2}
\newcommand{\RetX}[2]{\stcmd{s(#1)} & \quad #2 \\}

\section{The \mono{require} command}

\subsection{Description}

The \stcmd{require} command verifies if community-contributed packages installed on a system adhere to minimum or exact version requirements.
It can be used in both do-files and ado-files to specify code dependencies, ensuring they are met whenever the code is executed.


The command works in three steps. First, it finds the main ado-file associated with a package.
Second, it scans this ado-file for the so-called ``starbang lines'', special lines usually at the beginning of the file that start with the characters ``{\tt *!}'' (\dref{type}).
While starbang lines are not standardized, they typically contain the command name, its version, release date, and authors. See table \ref{table:starbang} for illustrative examples.
Third, it parses the starbang lines in order to extract the package version and release date, which can then be used to validate that any requirements are met.

\begin{table}[ht!]
\caption{Sample starbang lines of community-contributed packages}
\label{table:starbang}
\fontsize{10}{14}\selectfont
\begin{center}
\begin{tabular}{ll}
\hline
\noalign{\smallskip}
Package  & First starbang line                                          \\
\noalign{\smallskip}
\hline
ivreg2   & { *! ivreg2 4.1.11  22Nov2019                            } \\
psmatch2 & { *! version 4.0.12 30jan2016 E. Leuven, B. Sianesi      } \\
mdesc    & { *! mdesc Version 2.1 dan\_blanchette@unc.edu 25Aug2011 } \\
esttab   & { *! version 3.30  25mar2022  Ben Jann                   } \\
\noalign{\smallskip}
\hline
\end{tabular}
\end{center}
\end{table}

Note that, although starbang lines can be easily read using the \stcmd{which} command, their lack of standardization presents special challenges that have previously hindered practical package version control.
To work around this, \stcmd{require} is based on a battery of regular expressions that are able to parse most community-contributed commands.  There are many other corner cases outside of oddly-structured starbang lines. For instance, some packages distribute only Mata files; others distribute graphical styles; etc. For the most part, these are also handled by \stcmd{require}. 

Further, to simplify usage, \stcmd{require} standardizes version numbers and also extracts version release dates whenever available.\footnote{Standardized version numbers are based on semantic versioning, which defines software versions using the \texttt{major.minor.patch} format. Major updates (e.g. from version 1.1 to 2.0) represent updates that might break compatibility. Minor updates (from 1.1 to 1.2) correspond to new features. Patch updates (from 1.1 to 1.1.1) indicate bug fixes. For additional information, see \url{https://semver.org/}.}


\subsection{Syntax}

At its simplest, \stcmd{require} can be used to assert that a package is installed:

\begin{stsyntax}
require {\it package} \optional{, main\_options}
\end{stsyntax}

More usefully, it can be used to ensure an exact or minimum version:


\begin{stsyntax}
require {\it package} $==$ {\it version} \optional{, main\_options}
\end{stsyntax}
\begin{stsyntax}
require {\it package} $>=$ {\it version} \optional{, main\_options}
\end{stsyntax}

\begin{table}[ht!]
\label{table:main-options}
\fontsize{10}{14}\selectfont
\begin{center}
\begin{tabular}{ll}
\hline
\noalign{\smallskip}
\textit{main\_options}  & Description                                          \\
\noalign{\smallskip}
\hline
\texttt{install} & install package if needed \\
\texttt{from}(\textit{location}) & an URL, directory, or ``SSC'' (default) \\
\texttt{adopath}(\textit{dirname}) & install path \\
\noalign{\smallskip}
\hline
\end{tabular}
\end{center}
\end{table}

To avoid cluttering do-files with many ``\texttt{require \ldots}'' lines, users can also set multiple requirements in a single line:

\begin{stsyntax}
require {\it requirement} \optional{{\it requirement}} \ldots \, \optional{, main\_options}
\end{stsyntax}

Where {\it requirement} can be any of the three syntaxes outlined above: 

\begin{stsyntax}
{\it requirement} := \optional{{\it package} | {\it package} $==$ {\it version} | {\it package} $>=$ {\it version}}
\end{stsyntax}

For complex projects, however, our recommendation is to avoid long statements and instead list all requirements in a separate file:


\begin{stsyntax}
require using {\it filename} \optional{, main\_options}
\end{stsyntax}

Lastly, the \texttt{setup} option can be used to create a requirements file based on the currently installed packages:

\begin{stsyntax}
require \optional{using {\it filename}} , setup \optional{setup\_options}
\end{stsyntax}

\begin{table}[ht!]
\label{table:list-options}
\fontsize{10}{14}\selectfont
\begin{center}
\begin{tabular}{ll}
\hline
\noalign{\smallskip}
\textit{list\_options}  & Description                                          \\
\noalign{\smallskip}
\hline
\texttt{adopath}(\textit{dirname}) & path containing installed packages \\
\texttt{save} & equivalent to ``\texttt{require using requirements.txt \ldots}'' \\
\texttt{replace} & replace ``using'' file if it already exists \\
\texttt{minimum} & list minimum requirements (\texttt{=>}) instead of exact (\texttt{==}) \\
\texttt{stata} & add a line requiring the currently-installed Stata version \\
\noalign{\smallskip}
\hline
\end{tabular}
\end{center}
\end{table}

\subsection{Basic usage}


While \stcmd{require} can be utilized from the command line, its most common usage is 
at the beginning of do-files:

\begin{stverbatim}
\begin{verbatim}

--------- < mydofile.do > ---------
* Description of do-file
  clear all
  cls
  version 18
  require mdesc 2.1
  require estout >= 3.23
  ...
-----------------------------------
\end{verbatim}
\end{stverbatim}

The command will execute silently if its listed dependencies are met:

\begin{stlog}
. require regress >= 1.3
\nullskip
\end{stlog}

Conversely, if these dependencies are not satisfied, the command will halt execution with error code 2225:

\begin{stlog}
. cap noi require regress >= 9.8.7
require error: you are using version 1.3.3 of regress, but require at least version 9.8.7
{\smallskip}
\nullskip
\end{stlog}

\subsection{Structure of requirement files}

Requirement files list one dependency for each line, optionally specifying either a minimum or an exact version. You can specify alternative  installation sources through the \texttt{from()} option. Any line starting with ``$*$'' or ``$\#$'' is treated as a comment and subsequently ignored.
For instance, the example below combines all these three criteria into a valid requirements file:

\begin{stverbatim}
\begin{verbatim}
------- < requirements.txt > ------
* This is a comment
ivreg2 >= 4.1
esttab == 3.30
gtools, from(https://raw.githubusercontent.com/mcaceresb/stata-gtools/master/build/)
-----------------------------------
\end{verbatim}
\end{stverbatim}

This requirement can then be implemented within a do-file as follows:

\begin{stverbatim}
\begin{verbatim}
. require using requirements.txt, install
\end{verbatim}
\end{stverbatim}

Note that these files can have any name, but we encourage users to follow conventions from other programming languages by naming them \textit{requirements.txt}.

\subsection{Creating requirement files}

Crafting correct requirement files can be challenging, as it requires researchers to meticulously add all package dependencies as well as their exact or minimum versions. The \texttt{setup} option addresses this by displaying all the community-contributed packages currently installed in a format easily amenable for copy-pasting. This option also supports directly saving its output into a requirements file, further streamlining usage. By default, the \texttt{setup} option writes exact package version requirements ($==$), rather than minimum requirements ($>=$), which can be achieved with the \texttt{minimum} option.


\begin{stlog}
. require, setup
(listing installed packages based on file c:\\ado\\plus/stata.trk)
{\smallskip}
\oom
    moremata == 1.1.0
    rangestat == 1.1.1
    mdesc == 2.1.0
    xsvmat
\oom
{\smallskip}
\nullskip
\end{stlog}

\subsection{Saved results}
\texttt{require} saves the following results to \stcmd{s()}:

\begin{stresults2}
\stresultsgroup{Macros} \\
    \stcmd{s(package)} & \quad package name \\
    \stcmd{s(version)} & \quad version string of the form ``major.minor.patch'' \\
    \stcmd{s(version\_major)} & \quad major version \\
    \stcmd{s(version\_minor)} & \quad minor version \\
    \stcmd{s(version\_patch)} & \quad patch version \\
    \stcmd{s(version\_date)} & \quad release date in the \texttt{\%td} format (e.g. ``31dec2022") \\
    \stcmd{s(filename)} & \quad ado-file used to determine the version (typically \textit{package}.ado) \\
    \stcmd{s(raw\_line)} & \quad line of code within the ado-file used to determine the version
\end{stresults2}

\section{Examples}

\subsection{Validating dates}

For advanced applications of \stcmd{require}, users can leverage the results returned in {\tt s()}. As an illustration, the example below validates the \emph{version date} of the \stcmd{regress} command, ensuring that its release was not prior to the year 2020:

\begin{stlog}
. require regress
{\smallskip}
. sreturn list
{\smallskip}
macros:
       s(version_date) : "09dec2020"
            s(version) : "1.3.3"
      s(version_major) : "1"
      s(version_minor) : "3"
      s(version_patch) : "3"
           s(raw_line) : "*! version 1.3.3  09dec2020"
            s(full_fn) : "C:\\Bin\\Stata17\\ado\\base/r/regress.ado"
           s(filename) : "regress.ado"
            s(package) : "regress"
{\smallskip}
. local date = td(`s(version_date)')
{\smallskip}
. assert `date' >= td(1jan2020)
{\smallskip}
\nullskip
\end{stlog}

\subsection{Usage in larger research projects}

For complex or large research projects, we strongly recommend listing package requirements in text files. This approach prevents any duplication and verbosity that could otherwise arise, given that large projects often involve several do-files and depend on multiple community-contributed packages.

Below is an illustration of this approach. First, we create the requirements text file:

\begin{stverbatim}
\begin{verbatim}
------- < requirements.txt > ------
* Replication file for "Some Paper" by Someone (AER 2023)

* SSC dependencies:
  mdesc >= 0.9.4
  estout >= 3.23

* Github dependencies:
  gtools		>= 1.7.5	, from("https://github.com/mcaceresb/stata-gtools/raw/master/build/")
-----------------------------------
\end{verbatim}
\end{stverbatim}

Next, to ensure these requirements are satisfied, we add the corresponding line at the start of each do-file:

\begin{stverbatim}
\begin{verbatim}
require using requirements.txt, install
\end{verbatim}
\end{stverbatim}

Once this line is executed, all specified dependencies will be checked, and if not already present, will also be installed and made available within Stata. Note, however, that the \texttt{install} option should only be employed when all users collaborating on a script or otherwise executing it have consented to the possibility of installing or updating community-developed packages in their own systems.

\subsection{Ensuring reproducibility through a clean ado-path}

As a best practice, we encourage users to validate the completeness of their requirements file against a clean ado-path.
Doing so helps ensure that the requirements file contains all requisite package dependencies for a given project, and is particularly useful for researchers submitting code to journals in order to meet code and data availability policies.

The code below achieves this task. Note that changes to the ado-paths as shown below are temporary, and will revert upon exiting and reopening Stata.

\begin{stlog}
. * 1) Install dependencies into custom folder
. require using requirements.txt, install adopath("C:\\my-folder")
  ... require sjlatex , adopath(C:\\my-folder) from(http://www.stata-journal.com/production) install 
  ... require mdesc >= 0.9.4 , adopath(C:\\my-folder)  install 
  ... require estout >= 3.23 , adopath(C:\\my-folder)  install 
{\smallskip}
. 
. * 2) Set adopath to this custom folder
. cap adopath - SITE
{\smallskip}
. cap adopath - PLUS
{\smallskip}
. cap adopath - PERSONAL
{\smallskip}
. adopath ++ "C:\\my-folder"
  [1]              "C:\\my-folder"
  [2]  (BASE)      "C:\\Bin\\Stata17\\ado\\base/"
  [3]              "."
  [4]  (OLDPLACE)  "c:\\ado/"
{\smallskip}
. 
. * 3) Run project code
. do myproject.do
{\smallskip}
. sysuse auto
(1978 automobile data)
{\smallskip}
. mdesc
{\smallskip}
    Variable    {\VBAR}     Missing          Total     Percent Missing
\HLI{16}{\PLUS}\HLI{47}
           make {\VBAR}           0             74           0.00
          price {\VBAR}           0             74           0.00
            mpg {\VBAR}           0             74           0.00
          rep78 {\VBAR}           5             74           6.76
       headroom {\VBAR}           0             74           0.00
          trunk {\VBAR}           0             74           0.00
         weight {\VBAR}           0             74           0.00
         length {\VBAR}           0             74           0.00
           turn {\VBAR}           0             74           0.00
   displacement {\VBAR}           0             74           0.00
     gear_ratio {\VBAR}           0             74           0.00
        foreign {\VBAR}           0             74           0.00
\HLI{16}{\PLUS}\HLI{47}
{\smallskip}
. exit
{\smallskip}
end of do-file
\nullskip
\end{stlog}

In the example above, all listed dependencies are first installed into a custom folder.  Next, the ado-path is updated to this folder. This step helps prevent Stata from accessing previously installed packages unrelated to the requirements file.  The last line in the example runs the project code as a final step in the verification process. If the project code fails, the requirements file likely has missing or incompatible dependencies and should be corrected.\footnote{We encourage users to also explore the {\tt repado} command from the \stcmd{repkit} community-contributed package. This command succinctly deals with the ado-path modifications of step two of the example. Additionally, the \stcmd{packagesearch} community-contributed command can be used as the starting point for a requirements file. Given a project with corresponding do-files, \stcmd{packagesearch} searches these do-files for keywords matching existing community-contributed packages.}


\section{Potential limitations} 

We have identified three factors that might limit the usefulness and adoption of \stcmd{require}.
We discuss each of them below, including possible workarounds.

First, although \stcmd{require} can be used to install other packages, it cannot install or update itself if it is not already present in the system. In other words, one cannot ``\texttt{require require}.'' 
A simple solution to this problem involves including two additional lines of code before the call to \stcmd{require}:

\begin{stverbatim}
\begin{verbatim}
cap which require
if (c(rc)) net install require, ///
    from(https://raw.githubusercontent.com/sergiocorreia/stata-require/master/src/)
require using requirements.txt, install
\end{verbatim}
\end{stverbatim}

The first line verifies if the package is already installed, and the second installs it from Github if not.

A second limitation is that the SSC does not maintain an archive of historical package versions and instead only stores the most recent version of each package. Thus, it is not possible to install superseded package versions from the SSC.

There are two possible workarounds. At its simplest, researchers might opt to include their ado paths as part of their replication package, taking care to also incorporate any compiled files for operating systems other than their own---else, their code would not be reproducible on other operating systems.
A more holistic approach would involve creating a repository of historical SSC packages, using as inputs the SSC Mirror maintained by the \href{https://github.com/labordynamicsinstitute/ssc-mirror/}{Labor Dynamic Institute} since 2021, as well as the partial backup maintained by the \href{https://archive.org/}{Internet Archive} since 2013.

Third, \stcmd{require} might be less useful for researchers that rely on packages with versions that cannot be parsed by \stcmd{require}.
This is particularly critical given the long tail of community-contributed packages, and given that package developers have historically defined version strings in an idiosyncratic manner.
To tackle this, we built \stcmd{require} by testing it against the universe of existing SSC packages, as well as on the most popular Github packages.
We discuss the performance of this approach in depth in the next section.

\section{Performance}

\begin{figure}[!ht]
\begin{center}
\caption{Cumulative distribution of SSC package downloads. \label{fig:powerlaw}}
\includegraphics[width=0.8\textwidth]{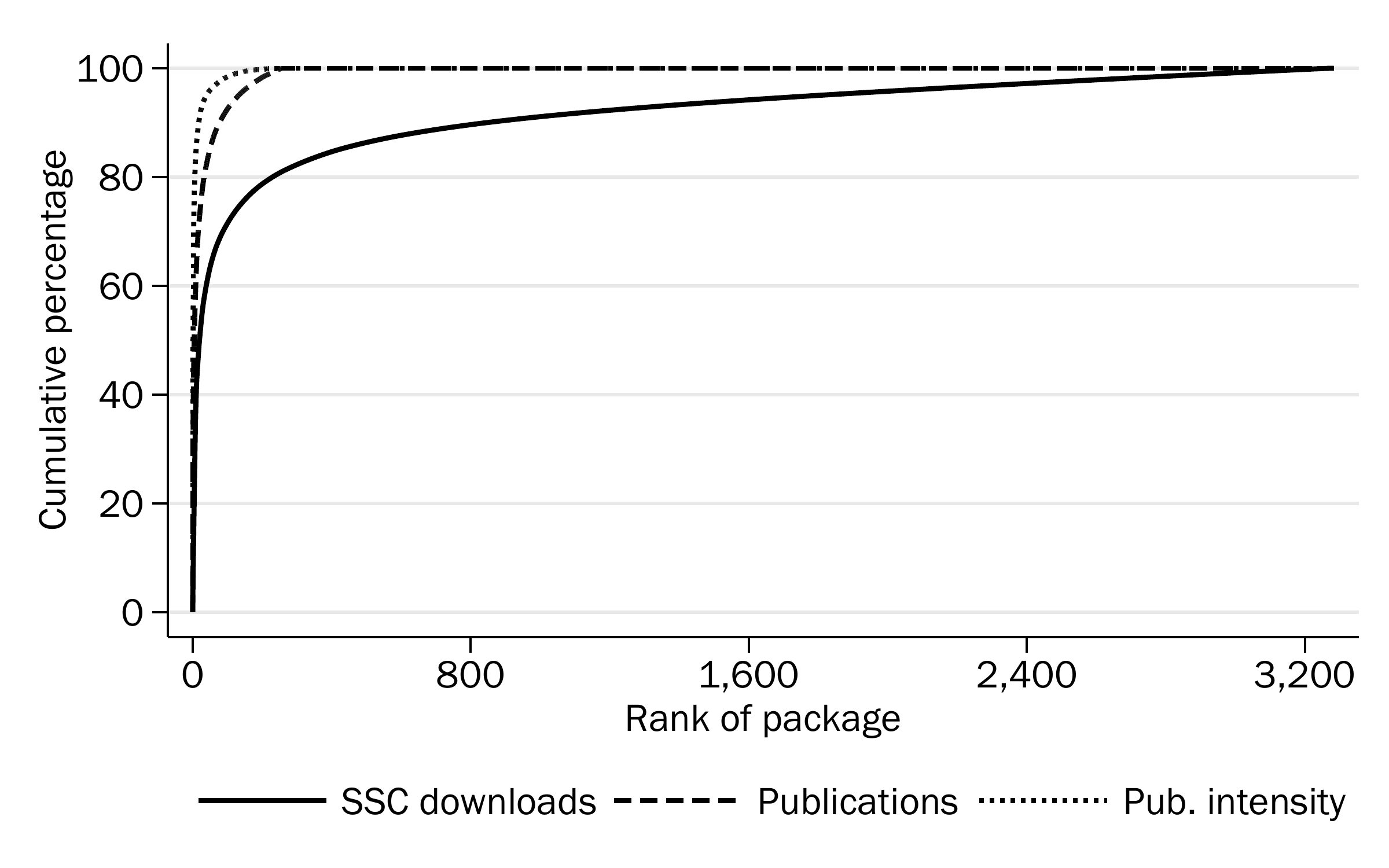}
\end{center}
\footnotesize{\textit{Notes}: Publication data based on analysis of journal replication by \citet{kranz2023}.}
\end{figure}

As of April 2024, there were roughly 3,500 packages available in the SSC. Their usage, however, is not uniform. For example, as shown in the solid line of figure \ref{fig:powerlaw}, the top 100 packages account for about 71 percent of all package downloads even though they correspond to about 3 percent of the SSC universe.

This skewness is even stronger when we replace package downloads with measures closer to actual usage by researchers.
Using data from \citet{kranz2023} we compile a list of package usage from the replication packages of publications in top economics journals.
We then compute the fraction of publications that rely on each community-contributed package (dashed line) as well as the ``publication intensity'' of each package---the total number of times a package is used across all publications (dotted line).
Using these metrics, package usage is even more skewed, with the top 100 packages accounting for 93 percent of all publications (green line) and 98 percent of all usage in code (red line).

\begin{figure}[!ht]
\begin{center}
\caption{Performance of \texttt{require} against SSC packages.\label{fig:performance}}
\includegraphics[width=0.8\textwidth]{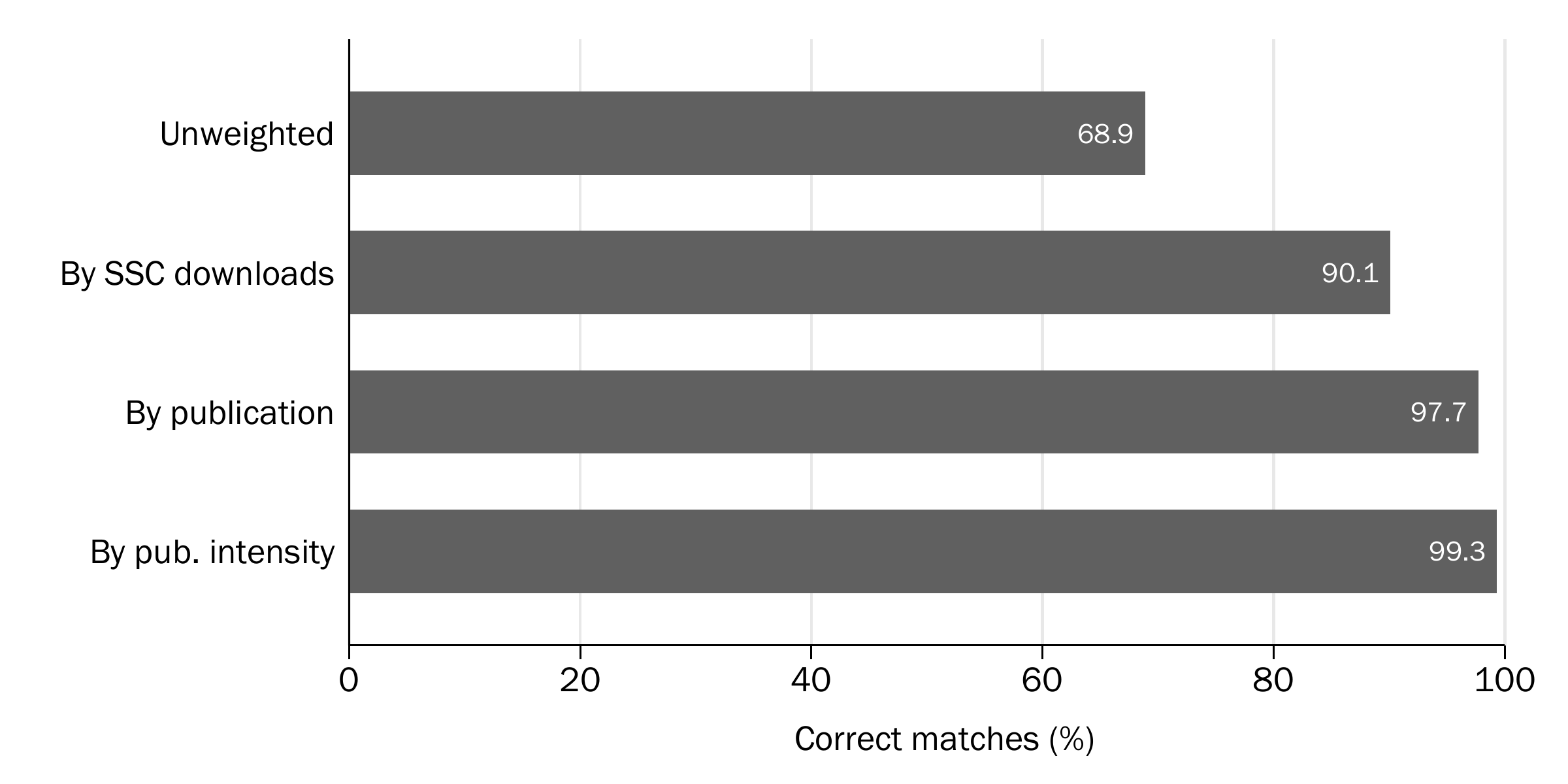}
\end{center}
\footnotesize{\textit{Notes}: Publication data based on analysis of journal replication by \citet{kranz2023}. }
\end{figure}

Given this distribution, \stcmd{require} relies on a hybrid approach. On the one hand, it uses a battery of patterns to try to match as many packages as possible. On the other, it has custom rules for popular packages whose starbang lines do not match these patterns.

To evaluate how the code performs, we first construct a manual ground truth dataset---version numbers based on a 2023 mirror of SSC---and then evaluate \stcmd{require} against this dataset. Figure \ref{fig:performance} shows the results.
The first bar shows that \stcmd{require} is able to extract version information for 69 percent of all SSC packages which is reasonable considering that the majority of the remaining packages lack a version string altogether.
The second bar shows that once we weight by the number of SSC downloads, the success rate of \stcmd{require} jumps to 90 percent.
Results are even stronger when considering publications.
Here, \stcmd{require} is able to match about 98 percent of all packages used in publications, and about 99 percent when weighted by publication intensity.

\section{Conclusions}

In this article, we introduced the \stcmd{require} command, which helps facilitate reproducible research in Stata by automating version control for community-contributed packages across co-authors, computing environments, and through time.  
The command correctly extracts the versions of most packages distributed on SSC, particularly those most widely used by researchers. 

While \stcmd{require} takes a small step towards increasing the reproducibility of research done using Stata, much work remains. The coverage offered by  \stcmd{require} would expand significantly if the SSC transitions towards a versioned package setup, or if additional efforts are taken to construct a database of Stata packages by version and by date. This step would enable users to reproduce a broad swathe of research built on prior package versions and would help build transparency in research findings, and add to our understanding of why inconsistencies may arise between published results and results using the latest Stata package versions.    

In addition, establishing a uniform syntax for declaring package versions would increase the fidelity of \stcmd{require} through time and help limit the long tail of unmatched packages that would otherwise require manual construction.

\bibliographystyle{sj}
\bibliography{literature}

\clearpage
\end{document}